\pdfoutput=1

\documentclass[11pt]{article}

\usepackage{ACL2023}

\usepackage{times}
\usepackage{latexsym}
\usepackage[T1]{fontenc}
\usepackage[utf8]{inputenc}
\usepackage{microtype}
\usepackage{tikz}

\usepackage{graphicx}
\usepackage{subcaption}
\usepackage{inconsolata}
\usepackage{enumitem}
\usepackage{booktabs}

\usepackage{comment}

\usepackage{amsmath}
\usepackage{amssymb}

\usepackage{cleveref}

\title{CafGa: Customizing Feature Attributions to Explain Language Models}

\author{
Alan Boyle \qquad
Furui Cheng\textsuperscript{*}\qquad 
Vilém Zouhar \qquad
Mennatallah El-Assady \\
ETH Zurich
}

\begin{document}

\maketitle

\begin{abstract}
Feature attribution methods, such as SHAP and LIME, explain machine learning model predictions by quantifying the influence of each input component. 
When applying feature attributions to explain language models, a basic question is defining the interpretable components.
Traditional feature attribution methods, commonly treat individual words as atomic units.
This is highly computationally inefficient for long-form text and fails to capture semantic information that spans multiple words.
To address this, we present CafGa, an interactive tool for generating and evaluating feature attribution explanations at customizable granularities. 
CafGa supports customized segmentation with user interaction and visualizes the deletion and insertion curves for explanation assessments. 
Through a user study involving participants of various expertise, we confirm CafGa's usefulness, particularly among LLM practitioners. 
Explanations created using CafGa were also perceived as more useful compared to those generated by two fully automatic baseline methods: PartitionSHAP and MExGen, suggesting the effectiveness of the system.
\end{abstract}

\renewcommand{\thefootnote}{\fnsymbol{footnote}}
\footnotetext[1]{Corresponding author}
\footnotetext[2]{CafGa is available as a pip-installable package \href{https://pypi.org/project/cafga}{cafga}. The source code is hosted on \href{https://github.com/explain-llm/CafGa}{https://github.com/explain-llm/CafGa}, and a live demo is accessible at \href{https://cafga.ivia.ch/}{cafga.ivia.ch}.}

\smallskip
\section{Introduction}
\label{sec:introduction}

Large Language Models (LLMs) continuously evolve in scale and capabilities across a wide range of tasks, such as question answering, reasoning, and text summarization \citep[inter alia]{kamalloo-etal-2023-evaluating}.
Meanwhile, their increasing complexity poses significant challenges in interpretability and human trust \citep{trustllm}. 
Feature attribution methods, e.g. SHAP \citep{shap} and LIME \citep{lime}, offer a promising approach to explain LLM behaviors by quantifying the influences of each component in the input.
 These explanations calibrate users' trust in model predictions \citep{10.1145/3411764.3445717} and offer useful insights into the prediction shortcuts and model biases \citep{du-etal-2021-towards,ren-xiong-2023-huaslim}.
\begin{figure}[t]
    \includegraphics[width=\linewidth]{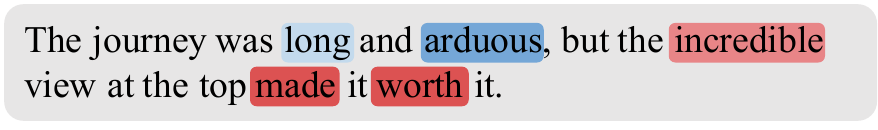}
    \caption{Feature attribution explanations quantify the contributions of each component (e.g., words), allowing users to validate the models' reasoning. }
    \label{fig:shap_intro}
\end{figure}
Defining interpretable components is a fundamental question in generating meaningful feature attributions~\citep{lime}. 
Traditional removal-based methods typically treat individual words as the basic units of interpretation (\Cref{fig:shap_intro}), but this approach struggles to scale effectively for long-form text due to computational inefficiency and the fact that meaningful semantic cues often span multiple tokens or phrases, such as \textit{``made it worth it''} in \Cref{fig:shap_intro}.
To address this, we propose CafGa, an interactive tool that enables users to create feature attribution explanations at customized levels of granularity. 
The tool offers default options to segment text at the word, sentence, or paragraph level, and supports users to further refine these segments by interactively isolating text spans.
CafGa also provides functionality to evaluate and compare different segmentation strategies.
The system visualizes deletion and insertion curves, showing how the model's prediction changes as components are removed or added based on their attribution rankings, and calculates the area under the perturbation curve as the fidelity scores.

We conducted a two-stage user study to evaluate the usability and effectiveness of CafGa. 
In the first phase, ten participants used CafGa to create explanations that helped them understand the model’s reasoning. 
Based on self-reported usability scores, users with machine learning expertise generally found the system easy to use and learn, while novice users experienced some difficulty due to the need to grasp new concepts before effectively interacting with the tool. 
Despite this, participants across all expertise levels reported that CafGa was helpful in improving their understanding of the LLM.
In the second phase, we invited four expert users to compare the usefulness of the CafGa-generated explanations from the first phase with those produced by two automated baselines: PartitionSHAP \citep{partitionshap} and MExGen \citep{MonteiroPaes2024MultiLevelEF}. 
In 64\% of the comparisons, participants rated the human-generated explanations as the most helpful, demonstrating the effectiveness of the proposed system in supporting meaningful model interpretation.

\begin{figure}[!ht]
    \centering
    \includegraphics[width=\linewidth]{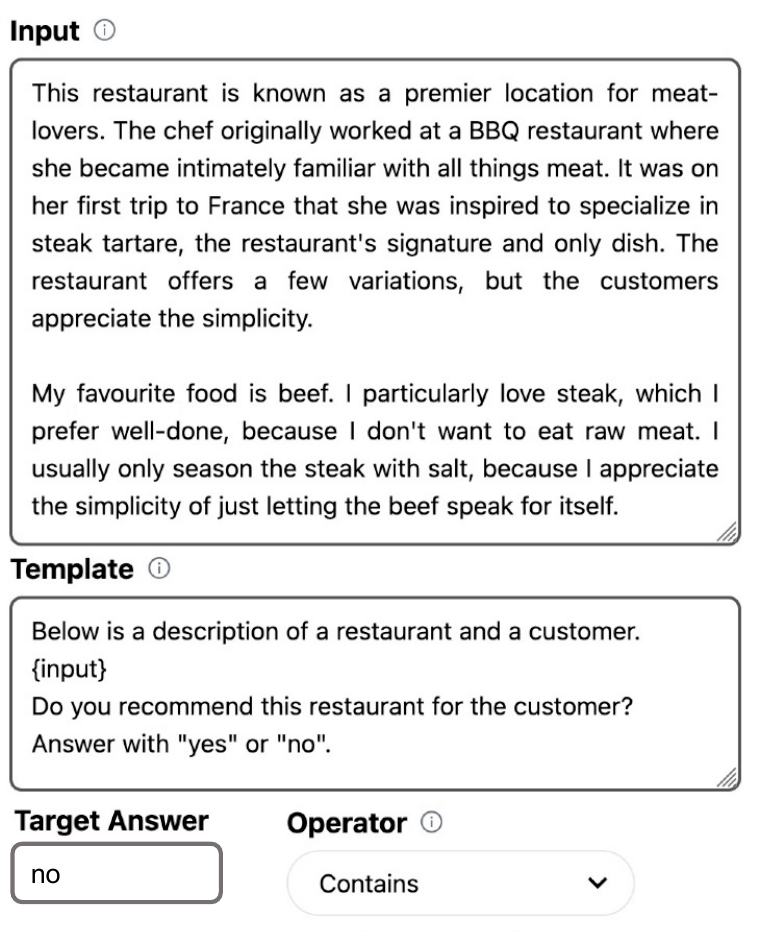}
    \caption{The user creates a task where a model is asked to decide whether to recommend a restaurant to a customer.}
    \label{fig:CS-Task}
\end{figure}

\section{System Design}
\label{sec:sys_design}

CafGa is an interactive system that includes a \textit{task creation page} for defining prediction tasks and customizing explanation granularities, and an \textit{explanation page} that displays explanations and their evaluation results. 
The following sections introduce the key features and design.

\begin{figure*}[t]
    \centering
    \includegraphics[width=\linewidth]{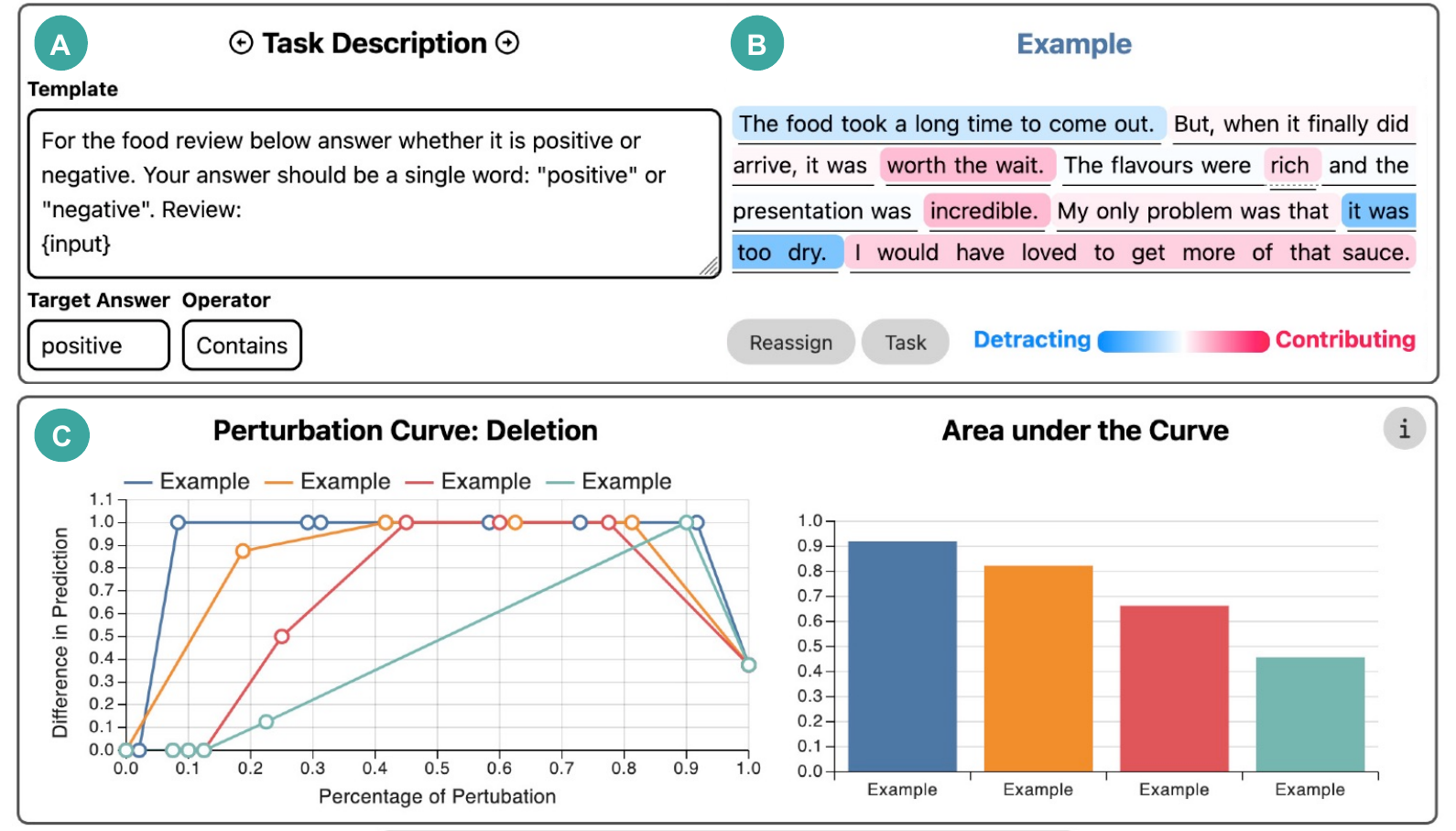}
    \caption{An overview of CafGa: The user first creates a predictive task (A) and then decides the granularities for explanations by assigning words into groups and gets an explanation based on the assignments (B). Using the perturbation curve the user can validate the fidelity of the generated explanation (C).}
    \label{fig:Fig1}

    \vspace{-4mm}
\end{figure*}

\paragraph{Define a Prediction Task.} 
CafGa allows users to define a prediction task by constructing a prompt and defining an evaluator (\Cref{fig:CS-Task}). 
To construct the prompt, users need to complete the \texttt{template} field with the structure of the prompt and the \texttt{input} field with the content they wish to explain. 
We use the design in ChainForge \citep{chainforge}
where an evaluator is used to structure the text generated by the model by transforming the text into a boolean value based on user-defined rules. 
To define the evaluator, users can choose from predefined operators such as \texttt{Contains}, which returns a boolean value indicating whether the text includes the specified \texttt{Target Answer}, and \texttt{Entails}, which assesses whether the text logically entails the \texttt{Target Answer}.
See the full list of operators in \Cref{sec:operator-desc}. 


\begin{figure}[htbp]
    \centering
    \includegraphics[width=\linewidth]{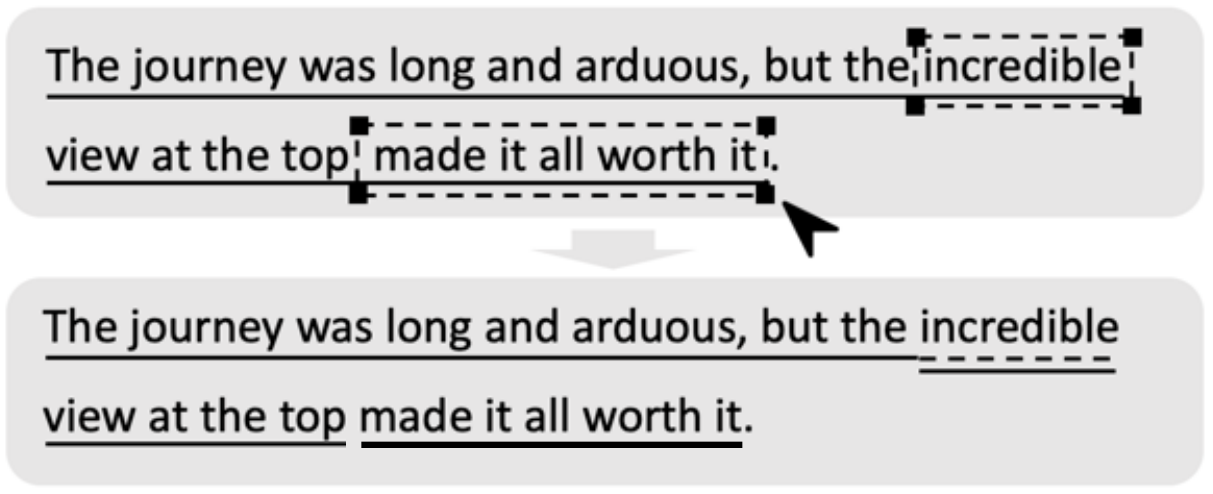}

    \vspace{-1mm}
    \caption{CafGa allows users to interactively segment text to customize the explanation granularity. }
    \label{fig:edit-page}

    \vspace{-3mm}
\end{figure}

\paragraph{Customize Text Segmentation.}
CafGa supports default options for creating text segments, such as dividing text by word, sentence, or paragraph. 
Users can customize the segmentation through interactions. 
For example, they could brush to select a phrase and isolate the selected phrase from the sentence (\Cref{fig:edit-page}). The segments may not overlap to avoid ambiguous attributions.

\paragraph{Calculate and View Feature Attributions.}
CafGa uses the KernelSHAP \cite{shap} algorithm to sample and compute the feature attributions. 
The system generates random samples by removing text segments in the user-specified granularities. 
For each perturbed sample, the system requests ten responses, which are converted into Boolean values using the user-defined evaluator.
The estimated probability, e.g., $\texttt{P(the answer contains ``no'')}$ is computed as the proportion of responses evaluated as true.
Finally, the system uses a weighted linear regression model to estimate the Shapley values of each group. 
This step runs in the back-end and is hidden from users. 
After the calculation, we visualize the explanations using a heatmap with a color-blind friendly color schema (\Cref{fig:Fig1}B) accompanied by the task description (\Cref{fig:Fig1}A). 

\paragraph{Evaluate Explanations.}
To ensure that the created explanations accurately reflect the model's decision making and mitigate users' confirmation bias, we adapt a commonly used and general fidelity metric, Area Over the Perturbation Curve \citep{VisualizingCNN_AOPC}.
Following \citet{petsiuk2018rise}, we employ two variants of the perturbation curve: \textit{Deletion} and \textit{Insertion}. 
Deletion begins with all the features present and then iteratively removes the highest valued features. Insertion begins with no features present and iteratively inserts the highest valued features.
In the resulting graph the x-axis represents the percentage of words added or removed and the y-axis the difference between the original prediction $f(x)$ and the perturbed prediction $f(x^{k})$ when perturbing the top $k$ features. Importantly, we plot the percentage of \textit{words} perturbed on the x-axis rather than the percentage of \textit{groups} to avoid a bias towards large groupings.
For deletion the area under the curve should be large as the perturbed prediction should quickly diverge from the original one. For insertion it should be small as the perturbed prediction should quickly converge back to the original one.
These metrics are visualized in an intuitive way to users such that they can follow and compare the computation exactly (\Cref{fig:Fig1}C).

\paragraph{Implementation Details.}
CafGa is implemented as a web application composed of a front-end and a back-end communicating via HTTP requests. The front-end is written in TypeScript and \href{https://react.dev/}{React} and served via \href{https://vitejs.dev/}{Vite}. 
The back-end server uses \href{https://fastapi.tiangolo.com/}{FastAPI} and \href{https://www.uvicorn.org/}{Uvicorn}. The model used in the back-end can be chosen by the user from a selection of \href{https://platform.openai.com/docs/models}{OpenAI models}. 
In the implementation of the KernelSHAP algorithm, we use the sampling strategy and functions from the \href{https://shap.readthedocs.io/en/latest/generated/shap.KernelExplainer.html#shap.KernelExplainer}{SHAP library}. This implementation uses $2 * n_{features} + 2048$ as the maximum number of samples used to approximate the shapley value. In the interactive setting we prioritize user experience and thus set the number of samples so that the explanation can be provided in a reasonable time $t_{max}$. Given a rate of API requests $r_{API}$, the maximum number of samples is defined as $n_{samples} = t_{max} \cdot r_{API}$.
We provide a PyPI package that contains all of the algorithms used in the back-end. The package also comes with Jupyter widgets created using AnyWidget \citep{anywidget} that recreate parts of the front-end. 
A Jupyter Notebook example is shown in \Cref{sec:jupyter}. This in particular allows users to also run CafGa on their local models. 
\section{A Worked-out Example}

\begin{figure*}[t]
    \centering
    \includegraphics[width=\linewidth]{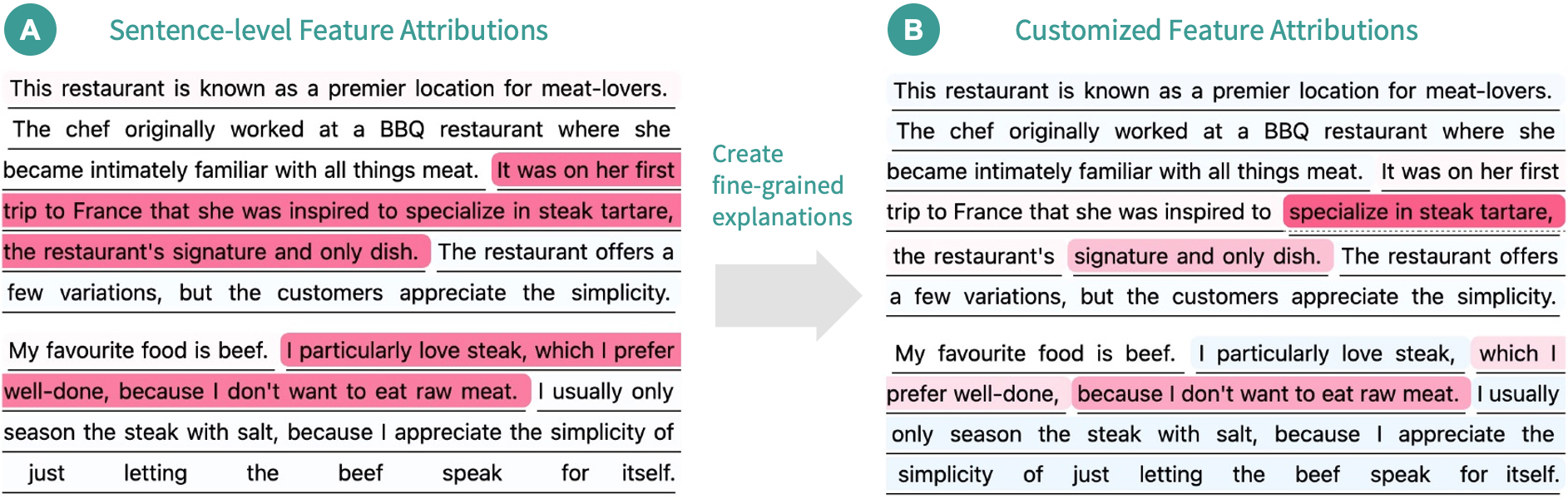}
    \caption{
    The user first used the sentence preset (A). The resulting explanation gives a rough overview of the model's reasoning.
    The user specifically assigned all relevant parts into separate groups (B). In (B), one can now clearly see the reason: tartare is the restaurant's sole signature dish, but the customer does not want to eat raw meat. 
    }
    \label{fig:CS-Steps}
\end{figure*}

We introduce a use case in understanding GPT4o-mini's predictions in multi-hop reasoning. 
In this task, the model needs to decide whether to recommend the restaurant to a customer based on the restaurant's review and the customer's preferences. 
The model makes a correct prediction that does not recommend the restaurant. 
The user, an LLM developer, wanted to understand if the model followed the correct reasoning behind this decision. 

The user first defined the task (see \Cref{fig:CS-Task}) and used a preset that calculates sentence-level attributions.
From the results (see \Cref{fig:CS-Steps} A), the user noticed that the two sentences with the highest attribution include the necessary information for making this decision -- \textit{the restaurant only serves steak tartare}, while \textit{the customer doesn't like raw meat} (see highlighted parts in \Cref{fig:CS-Steps} B). 
The user found this explanation unsatisfactory and wanted to know if the model specifically captured the key information that \textit{steak tartare} is \textit{the only dish served} in these sentences. 

So the user then manually separated the sentence into segments, like \textit{specialize in steak tartare}, \textit{signature and only dish}, to get a more precise explanation.  
The new explanation, with a higher fidelity (from 0.77 to 0.96) suggests that all the necessary information is well-captured by the model and has high contributions to the prediction. 
From this two-stage exploration, the user confirmed that the model follows the correct reasoning in this decision. 

\section{User Study}

To evaluate the usability of CafGa, we conducted a two-stage user study in which participants created and assessed explanations. In the first stage, participants used CafGa to generate customized explanations and then provided feedback on the tool’s usability. In the second stage, experts compared these participant-generated explanations with those produced by existing fully automatic methods to assess the quality of the generated explanations.

\subsection{Setup}

We now describe the two parts of the user study.

\paragraph{Creating explanations.}

For the first stage we recruited ten participants: six users who had experience working with machine learning, noted as experts and four novice users.
None of the participants had any prior experience working with attribution-based explanations.

At the beginning of the study, the participants were shown a video that explained the concepts behind CafGa and their role in this part of the study.
The video also showed an example task that the participants could work through while following the video.
After viewing the video participants were encouraged to ask questions.
The participants then created explanations for five types of tasks of increasing complexity.
The tasks were local question answering taken from SQuAD \citep{Squad}, sentiment analysis on reviews taken from the YELP academic dataset \citep{YELP}, a few-shot prompt engineering task inspired by \citet{tenney2024interactivepromptdebuggingsequence}, multi-hop reasoning taken from HotpotQA \citep{hotpotqa} and long-form text comprehension taken from BARQA \citep{hou-2020-bridging} (see Appendix \ref{sec:app-task-desc} for details).
For each task type there were five tasks except for prompt-engineering and long-form text comprehension, which each had three tasks.
The given task for each type was selected at random.
Once participants had created an explanation for each type of task, they filled out a survey rating the usability and usefulness of CafGa.

\paragraph{Comparing explanations.}
In the second stage, we recruited four expert participants who all studied machine learning and currently work with AI.
The participants were again shown a video that explained the concepts behind CafGa and their role in this part of the study.
The participants were asked to choose the most helpful explanations among 10 groups. 
Each group contained three feature attributions: a human-created explanation from the first stage of the study, one created by MExGen \citep{MonteiroPaes2024MultiLevelEF}, and one created by PartitionSHAP \citep{partitionshap}. We choose MExGen because it provides a good baseline of what static granularities can achieve and PartitionSHAP because it is often cited as a comparison \citep{amara-etal-2024-syntaxshap, MonteiroPaes2024MultiLevelEF, mosca-etal-2022-shapsurvey} and is recommended as an efficient method by \citet{mosca-etal-2022-shapsurvey}. Unlike most methods they also scale to long inputs like those in BARQA.
Participants were asked to select the most helpful explanation -- the one that best informed them of the model's decision logic and correctness -- among the three.

\subsection{Study Results} 

From the usability survey results (\Cref{tab:Survey-Results}), we observed that the experts could use the system well.
They generally found the system easy to learn and easy to use. 
In comparison, non-expert users faced more challenges in learning the system, primarily due to difficulties in quickly understanding the underlying concepts of CafGa.
However, after getting familiar with the system, both groups of participants reported that the system is helpful in supporting them in understanding the LLM.

From the comparative study results (\Cref{tab:Comparison-Results}), we find that the explanations made by humans were generally preferred, with MExGen outperforming PartitionSHAP. We see that in total, the explanations created by humans were preferred 64\% of the time over MExGen and PartitionSHAP. 
This demonstrates that CafGa is effective in supporting meaningful model interpretation. 

Task-level preferences (\Cref{tab:Comparison-Results}) show that human-customized explanations performed especially well on the HotpotQA task. This is likely because the input text in these tasks typically contains multiple facts, often expressed through distinct clauses and phrases.
The participants can make semantic segmentations to the input text, while automatic methods create unreasonable n-grams that do not align with human intuitions.

\newcommand{\customBarChart}[7]{
    \begin{tikzpicture}[x=0.3em, y=0.24em]
        \draw[red] (3.5, 0) -- (3.5, 4);
        \draw[use as bounding box] (0,0) rectangle (7, 0);
        \fill[fill=black]  (0,0) rectangle (1, #1);
        \fill[fill=black]  (1,0) rectangle (2, #2);
        \fill[fill=black]  (2,0) rectangle (3, #3);
        \fill[fill=black]  (3,0) rectangle (4, #4);
        \fill[fill=black]  (4,0) rectangle (5, #5);
        \fill[fill=black]  (5,0) rectangle (6, #6);
        \fill[fill=black]  (6,0) rectangle (7, #7);
    \end{tikzpicture}
}

\begin{table}[t]
    \small
    \centering
    \renewcommand{\arraystretch}{1.5}
    \begin{tabular}{p{3.8cm}cc}
    \toprule
    \bf Statement & \hspace{-5mm} \bf Non-experts & \bf Experts \\
    \midrule
    The system is easy to use.
    & 3.0 \customBarChart{0}{1}{2}{1}{0}{0}{0}
    & 5.0 \customBarChart{0}{0}{1}{0}{3}{2}{0} \\
    The system is easy to learn.
    & 3.5 \customBarChart{0}{0}{3}{0}{1}{0}{0}
    & 5.2 \customBarChart{0}{0}{0}{1}{4}{0}{1} \\
    The system is enjoyable to use.
    & 3.5 \customBarChart{0}{1}{1}{1}{1}{0}{0}
    & 5.8 \customBarChart{0}{0}{0}{0}{1}{5}{0} \\
    The various functions in the system are well integrated.
    & 5.0 \customBarChart{0}{0}{0}{2}{0}{2}{0}
    & 5.5 \customBarChart{0}{0}{1}{0}{2}{2}{1} \\
    The system is helpful in understanding the LLM.
    & 6.3 \customBarChart{0}{0}{0}{0}{0}{3}{1}
    & 5.8 \customBarChart{0}{0}{0}{0}{3}{1}{2} \\
    I would like to use this system frequently.
    & 3.3 \customBarChart{1}{0}{1}{1}{1}{0}{0}
    & 4.0 \customBarChart{1}{0}{0}{3}{0}{2}{0} \\
    \bottomrule
    \end{tabular}
    \centering
    \caption{After-task survey results. Participants were asked to rate the statements on a scale from 1 (strongly disagree) to 7 (strongly agree) with 4 (neutral). For each statement, we show the distribution and average.}
    \label{tab:Survey-Results}
\end{table}

\begin{table}
    \centering
    \small
    \begin{tabular}{rcc@{\hspace{2mm}}c}
    \toprule
    & \bf Human
    & \bf MExGen
    & \bf PartitionSHAP \\
    \midrule
    SQuAD & 63\% & 38\% & \phantom{0}0\%\\
    YELP & 67\% & 33\% & \phantom{0}0\% \\
    Prompt & 50\% & \phantom{0}0\% & 50\% \\
    HotpotQA & 75\% & 25\% & \phantom{0}0\% \\
    BARQA & 63\% & 38\% & \phantom{0}0\% \\[0.5em]
    Average 
    & 64\% & 28\% & \phantom{0}8\% \\
    \bottomrule
    \end{tabular}
    \caption{Preference win-rate for generated explanations across tasks.}
    \label{tab:Comparison-Results}
\end{table}

\section{Related Work}
\label{sec:relwork}

In this section we discuss related work on creating feature attributions at interpretable granularities and existing user interfaces.

\subsection{Interpretable Feature Attributions}
\label{sec:AttMethods}

Recent work in attribution-based explanations has focused on creating attributions at granularities beyond the word level. This approach can not only improve computational efficiency, but also make for more interpretable explanations.

Early work in this direction focuses on finding specific decompositions of the neural network to get contextual attributions in addition to token-level attributions \citep{james2018CD,singh2018ACD,jin-2019-towards-hierarchical}.
However, given that there are now many settings where one does not have access to the model's internal structure, such model-specific methods may no longer be applicable. Thus, there has been a shift towards model-agnostic methods that do not rely on the model's internal structure.

\citet{chen-etal-2020-generating-hierarchical} present a model-agnostic approach that iteratively splits the input sequence into smaller groups generating an attribution for each group. This hierarchical approach in particular allows users to validate whether the model can reason about the compositional nature of language.

\citet{ju-etal-2023-hierarchical} further improve on this by addressing the limitation that the spans of grouped features must be contiguous.
This is important because modern NLP architectures can model long-range non-contiguous dependencies in the input.
However, both of these methods struggle to scale to large input sequences. 

\citet{MonteiroPaes2024MultiLevelEF} accordingly propose a method in which they first group the input at the coarsest granularity (e.g., paragraphs) and then iteratively refine the granularity only for the highest attributed group. 
This approach allows them to greatly reduce the computation time, but it can fail to precisely pinpoint the important parts of the input that are in the groups which do not get the highest attribution score. 
It also relies on the assumption that the fixed granularities used for the explanation are interpretable.

\citet{Colab} emphasize the importance of the human understanding of language in creating explanations. They thus group the features in the input by use of a syntactic tree allowing them to make semantically meaningful groups.

Despite research into improving explanations by attributing features at differing granularities, little work has been done to directly involve humans in the creation of such explanations.
We close this gap by empowering users to customize the attributions in a way that best suits their needs, while still informing them of the fidelity of the created explanations to stay aligned with the model.

\subsection{LLM Explanation Interfaces}
\label{sec:ExplanationInterfaces}

Recent research has focused on helping users gain a deeper understanding of LLMs through interactive visual interfaces.

Many of these interfaces target specific neural networks and explain the model by visualizing the architecture and the model's inner computations \citep{cnnvis,rnnvis,lstmvis,seq-2-seq-vis}. 

However, as it may not always be possible to access the model's internal structure, there has also been a rise in model-agnostic LLM interfaces \citep{KnowledgeVIS,cheng2024relic,chainforge,LLMComparator}.
Many of these tools allow users to investigate \textit{what} a model will answer, but they do not provide users with precise tools to explain \textit{why} a model answers that way.

LLM Checkup \citep{wang-etal-2024-llmcheckup} allows users to explore the model's reasoning by chatting with the model. This approach relies on model self-explanation, however, which can be misleading as the explanations often lack fidelity \citep{madsen-etal-2024-self,self-expl-2}. 

Closely related to our work, \citet{Colab} analyze the LLM's predictions by perturbing the input with meaningful counterfactuals.
They use a syntactic tree to define semantically meaningful segments in the input which can be ablated in a perturbation.
This allows users to perform an analysis similar to attribution-based explanation, as they can test which of the defined segments are relevant for the model's prediction.

\citet{tenney2024interactivepromptdebuggingsequence} also present an interface where users can choose the granularity of feature attribution. However, their tool relies on model-specific methods limiting its applicability and only allows users to specify the features at fixed granularities or by use of a regex, which means customizing explanations is largely not possible. 

We further emphasize the need for users to enabled to fully customize the explanation rather than just choosing from a limited set of options. Our model-agnostic approach also makes our system more broadly applicable and the use of the fidelity metric allows users to make certain that their custom explanations have high fidelity.
\section{Conclusion}

We introduce CafGa, a tool that enables users to generate explanations at arbitrary granularities by grouping words in a customized manner. We argue that using CafGa users can create more interpretable explanations. CafGa also provides users with a fidelity metric to ensure that the created explanations still align with the model's decision making. Through a case study and a user study, we demonstrate CafGa's usability and effectiveness, finding that users successfully created explanations that were preferred over automatic methods. This highlights the value of human involvement in creating interpretable explanations.

\section*{Limitations}

Our user study was limited by a small sample size due to the constrained number of available participants.
In future work, we would like to conduct a larger-scale evaluation. In particular, we believe that evaluating the practical use of CafGa in real-world, non-laboratory settings would provide valuable insights and offer a more accurate assessment of its usability and effectiveness.

Approximating the Shapley value involves a trade-off between speed and accuracy. Using fewer groups increases computational efficiency but risks obscuring important information within those groups. Conversely, creating many word groups improves fidelity but substantially reduces speed. 
Providing users with appropriate guidance in balancing this trade-off and thereby determining a suitable number of groups remains an important direction for future investigation.

We use deletion and insertion as fidelity metrics which are the most common fidelity metrics for attribution-based explanations and can be intuitively visualized for users. 
However, these metrics introduce a degree of circularity as both the explanation and the evaluation are based on perturbation. 
Their applicability also becomes questionable when features represent groups rather than atomic units (i.e., words). To mitigate this, we plot the results with respect to atomic units rather than grouped features. Nonetheless, further research is needed to develop more reliable fidelity metrics.

\hfill
\section*{Ethics Statement}

The participants were sourced from a pool of academic labs, which work on reciprocal basis instead of monetary rewards.


\clearpage
\appendix
\section{Jupyter Notebook Usage}
\label{sec:jupyter}

\begin{figure}[!h]
    \centering
    \includegraphics[width=\linewidth]{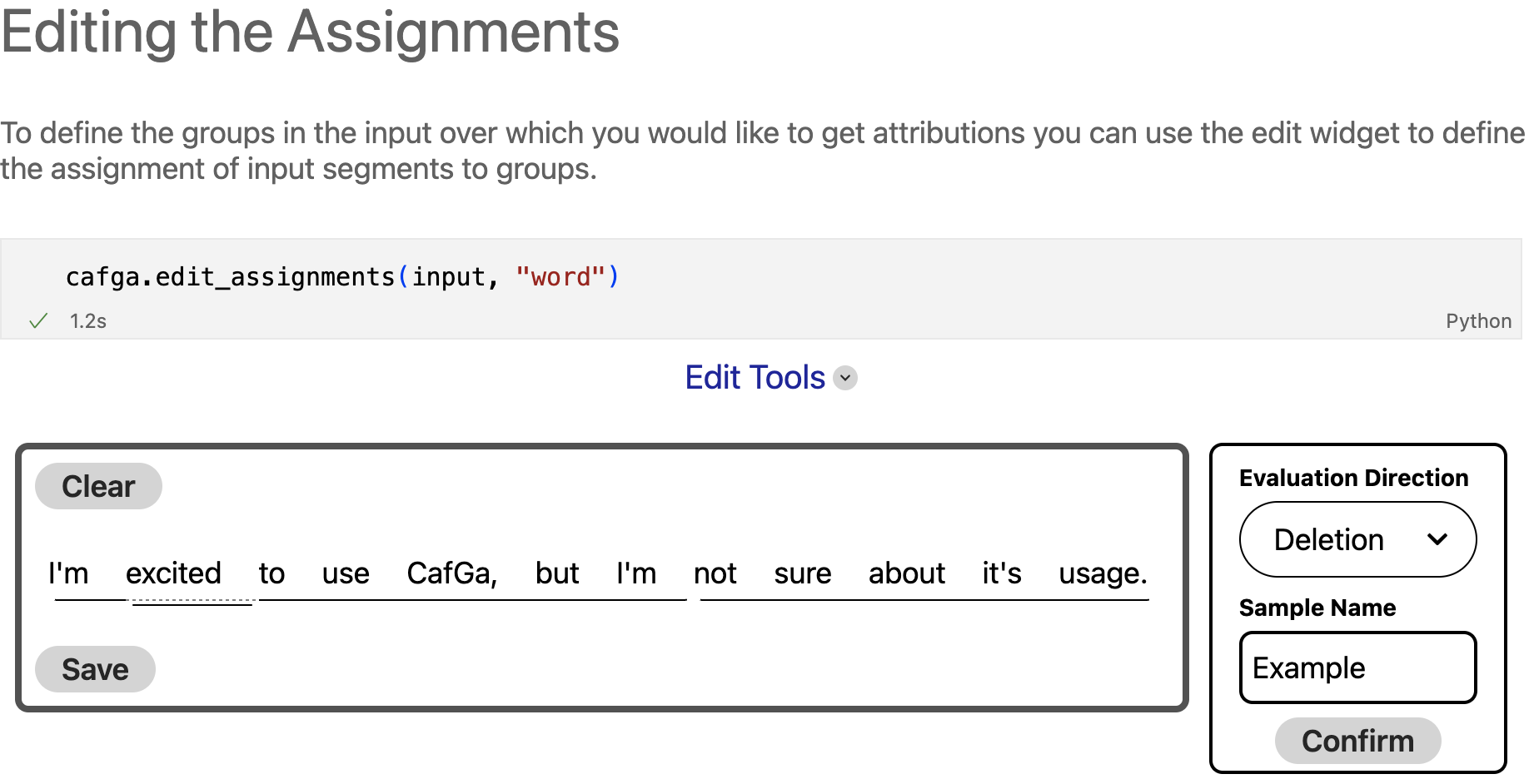}
    \includegraphics[width=\linewidth]{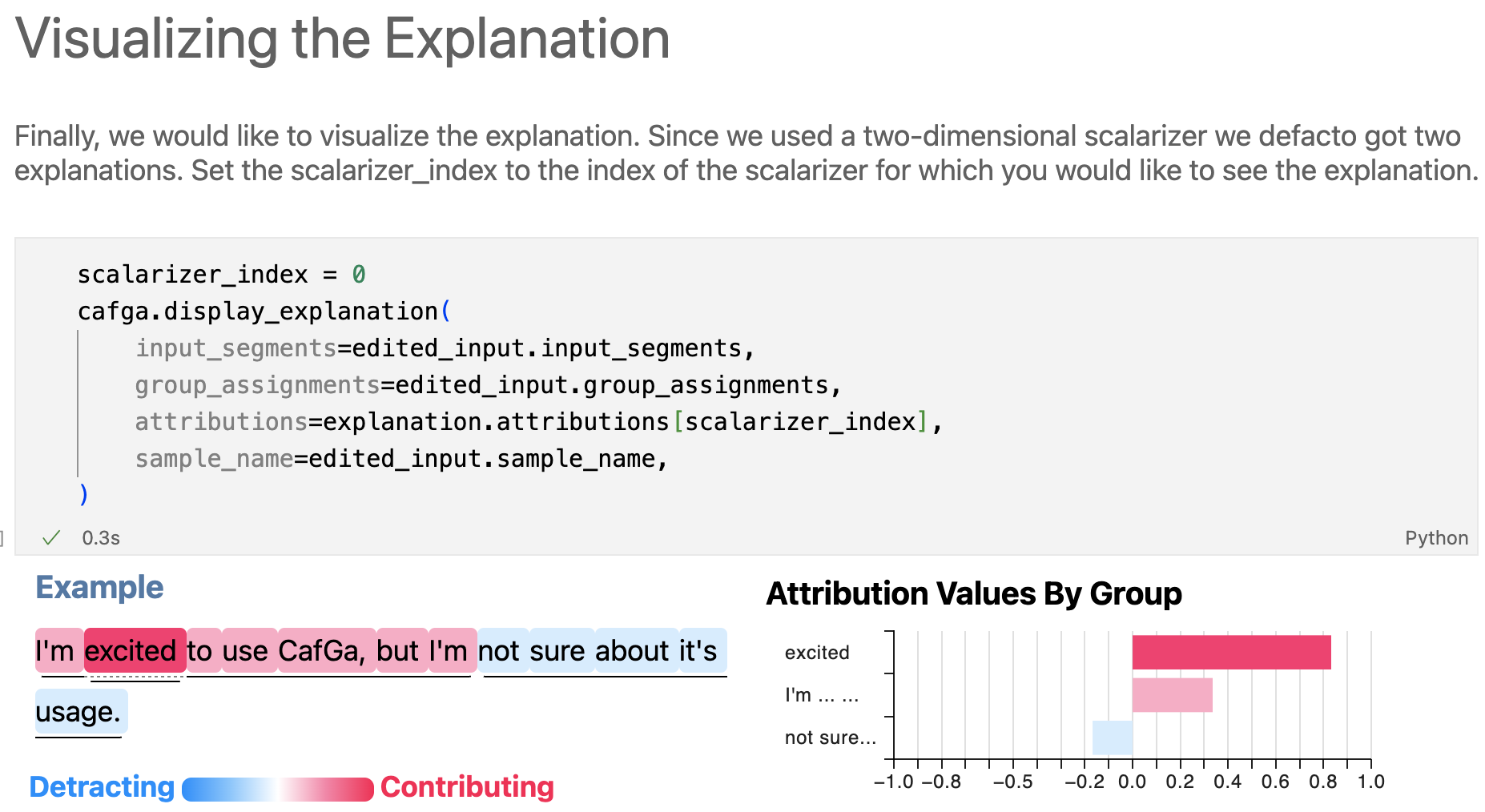}
    \caption{The assignment editor and the attributed text display as Jupyter notebook widgets.}
    \label{fig:widgets}
    
\end{figure}

\section{Operator Descriptions}
\label{sec:operator-desc}

The operators provided in CafGa can be seen in \Cref{tab:operators}. For the boolean operators we sample 10 model responses and use the percentage of times the operator is true to evaluate the models response. For the logical operators we sample a single response and use a DeBERTa-based NLI model \citep{nlimodel} to evaluate the response.

\begin{table}[h]
    \small
    \begin{tabular}{p{2cm} p{4.5cm}}
        \toprule
         \bf Name & \bf Description \\
         \midrule
         Contains & Checks whether the response contains the target answer \\
         Equals & Checks whether the response equals the target answer \\
         Starts With & Checks whether the response starts with the target answer\\
         Ends With & Checks whether the response ends with the target answer \\
         
         Entails & Checks whether the response logically entails the target answer \\
         Contradicts & Checks whether the response logically contradicts the target answer \\
         Semantically Equals & Checks whether the response semantically equals the target answer by applying entailment in both directions.\\
         \bottomrule
         
    \end{tabular}
    \centering
    \caption{The operators available in CafGa.}
    \label{tab:operators}
\end{table}
\section{Task Descriptions}
\label{sec:app-task-desc}

\textbf{SQuAD:} We take local question answering tasks from SQuAD \citep{Squad} and place the question in the template and the context in the input. The model is asked to answer the question given the context.\\
\textbf{YELP:} We take reviews from the YELP academic dataset \citep{YELP} and ask the model to predict whether the review is of positive or negative sentiment. We place the review in the input and the prediction instructions in the template.\\
\textbf{Prompt:} Inspired by the example presented in \citet{tenney2024interactivepromptdebuggingsequence} we create a set of few-shot prompts that contain errors. The explanation can be used to detect the errors and also to note parts in the instructions that cause the model to recreate the errors in the examples. We put the instructions and the examples in the input. The template only contains the new sample for which the user wants the model to follow the prompt.\\
\textbf{HotpotQA:} To create complex questions we use Hotpot QA \citet{hotpotqa}, which contains questions that require chaining multiple supporting facts. We put supporting facts, potentially misleading facts and the question in the input. The template only contains the instructions to answer the question.\\
\textbf{BARQA:} For long-form text comprehension we use examples from the Bridging Anaphora dataset \citet{hou-2020-bridging}. We place the article in the input and the question in the template.\\



\end{document}